\documentclass[sigconf]{acmart}
\usepackage{multirow}
\usepackage{multicol}
\usepackage{graphicx} 
\usepackage{graphicx}
\usepackage{caption}
\usepackage{subcaption}
\usepackage{algorithm}
\usepackage{algpseudocode}
\usepackage{soul}

\AtBeginDocument{%
  \providecommand\BibTeX{{%
    \normalfont B\kern-0.5em{\scshape i\kern-0.25em b}\kern-0.8em\TeX}}}

\setcopyright{acmlicensed}
\copyrightyear{2018}
\acmYear{2018}
\acmDOI{XXXXXXX.XXXXXXX}

\acmConference[Conference acronym 'XX]{Make sure to enter the correct
  conference title from your rights confirmation emai}{June 03--05,
  2018}{Woodstock, NY}
\acmISBN{978-1-4503-XXXX-X/18/06}

\begin{document}

\title{Automated Similarity Metric Generation for Recommendation}
\author{Liang Qu}
\authornote{Both authors contributed equally to this research.}
\email{liang.qu@uq.edu.au}
\affiliation{%
  \institution{The University of Queensland}
  \city{Brisbane}
  \country{Australia}
}

\author{Yun Lin}
\authornotemark[1]
\email{linyun@stu.pku.edu.cn}
\affiliation{%
  \institution{Peking University}
  \city{Beijing}
  \country{China}
}

\author{Wei Yuan}
\email{w.yuan@uq.edu.au}
\affiliation{%
  \institution{The University of Queensland}
  \city{Brisbane}
  \country{Australia}
}

\author{Xiaojun Wan}
\email{wanxiaojun@pku.edu.cn}
\affiliation{%
  \institution{Peking University}
  \city{Beijing}
  \country{China}
}

\author{Yuhui Shi}
\authornote{Corresponding Author}
\email{shiyh@sustech.edu.cn}
\affiliation{%
  \institution{Southern University of Science and Technology}
  \city{Shenzhen}
  \country{China}
}

\author{Hongzhi Yin}
\authornotemark[2]
\email{db.hongzhi@gmail.com}
\affiliation{%
  \institution{The University of Queensland}
  \city{Brisbane}
  \country{Australia}
}

\renewcommand{\shortauthors}{Trovato and Tobin, et al.}

\begin{abstract}
The embedding-based architecture has become the dominant approach in modern recommender systems, mapping users and items into a compact vector space. It then employs predefined similarity metrics, such as the inner product, to calculate similarity scores between user and item embeddings, thereby guiding the recommendation of items that align closely with a user's preferences. Given the critical role of similarity metrics in recommender systems, existing methods mainly employ handcrafted similarity metrics to capture the complex characteristics of user-item interactions. Yet, handcrafted metrics may not fully capture the diverse range of similarity patterns that can significantly vary across different domains.

To address this issue, we propose an Automated Similarity Metric Generation method for recommendations, named AutoSMG, which can generate tailored similarity metrics for various domains and datasets. Specifically, we first construct a similarity metric space by sampling from a set of basic embedding operators, which are then integrated into computational graphs to represent metrics. We employ an evolutionary algorithm to search for the optimal metrics within this metric space iteratively. To improve search efficiency, we utilize an early stopping strategy and a surrogate model to approximate the performance of candidate metrics instead of fully training models. Notably, our proposed method is model-agnostic, which can seamlessly plugin into different recommendation model architectures. The proposed method is validated on three public recommendation datasets across various domains in the Top-K recommendation task, and experimental results demonstrate that AutoSMG outperforms both commonly used handcrafted metrics and those generated by other search strategies.

\end{abstract}

\ccsdesc[500]{Information systems~Recommender systems}

\keywords{Recommender systems, Automated machine learning, Similarity metric, Evolutionary algorithm}


\maketitle

\section{Introduction}

\begin{figure}[htbp]
  \centering
  \begin{subfigure}[b]{0.35\linewidth}
    \includegraphics[width=\linewidth]{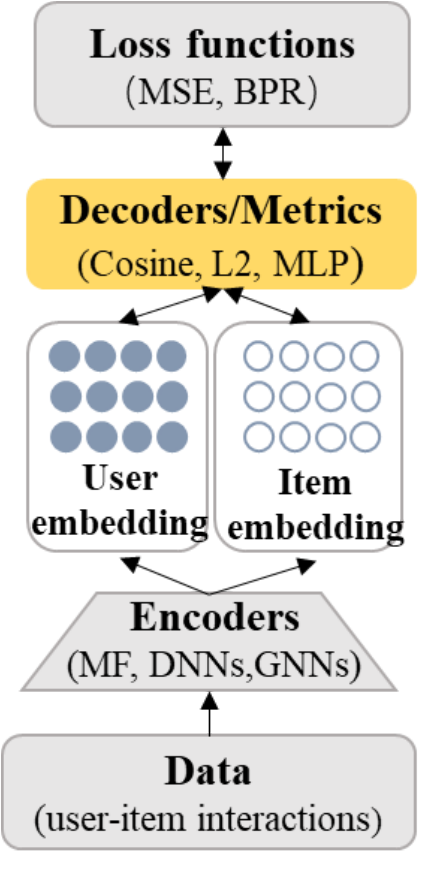}
    \caption{The architecture of RSs.}
    \label{fig:RSArchitecture}
  \end{subfigure}%
  \hspace{0.5mm}
  \begin{subfigure}[b]{0.6\linewidth}
    \includegraphics[width=\linewidth]{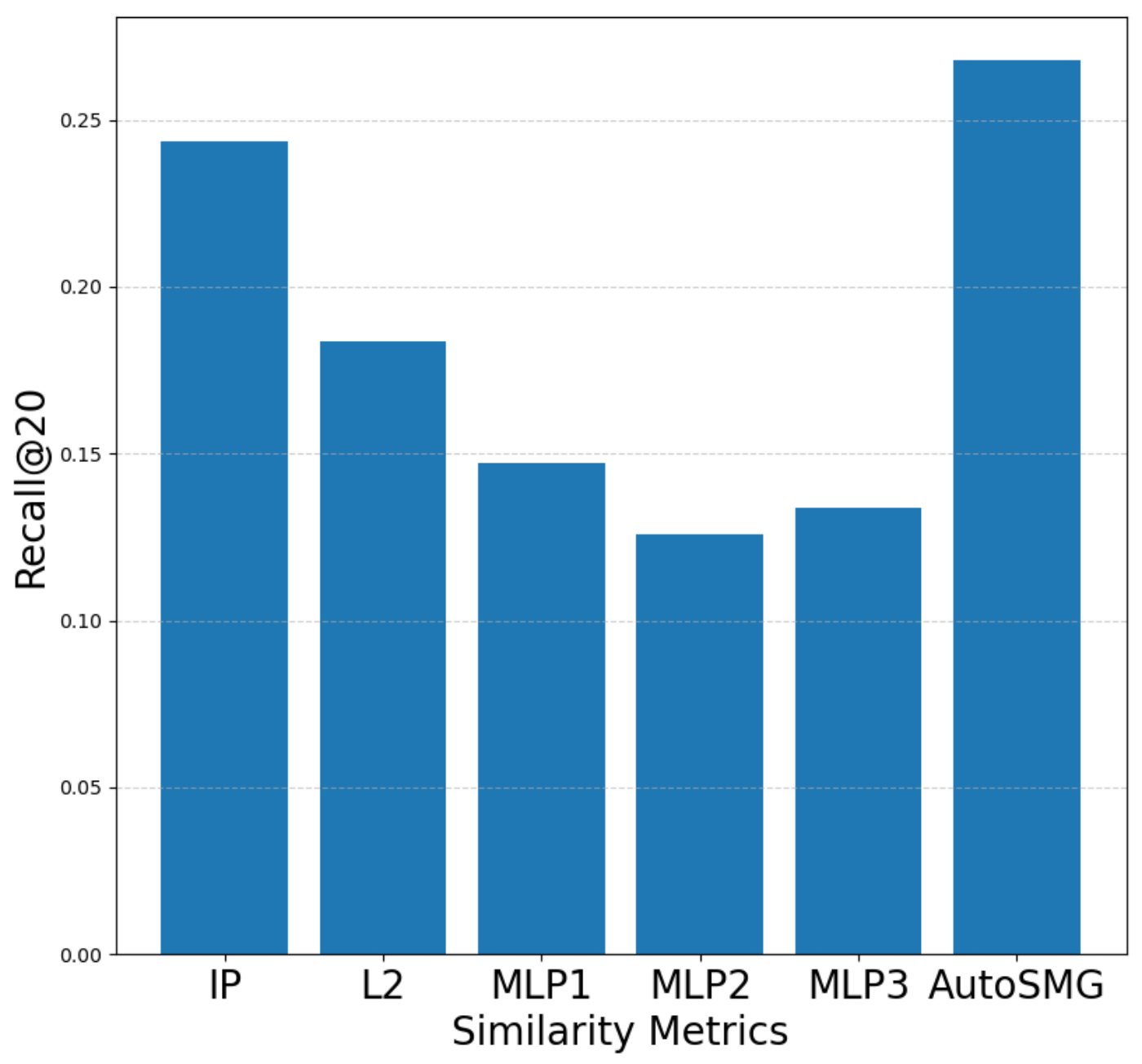}
    \caption{Performance of LightGCN \cite{he2020lightgcn} with different similarity metrics on MovieLens-1M dataset \cite{harper2015movielens}. IP denotes the inner product. MLP1, MLP2, and MLP3 represent various MLP architectures.}
    \label{fig:pe}
  \end{subfigure}
  \caption{Preliminary experiments.}

\end{figure}

Recommender systems (RSs) \cite{burke2002hybrid} are widely applied across various applications to assist users in finding options that match their preferences from a vast array of information.
The typical architecture of RSs can be summarized as an encoder-decoder framework \cite{mu2018survey}. As illustrated in Figure \ref{fig:RSArchitecture}, this framework first utilizes an encoder to learn representations (i.e., embeddings) of users and items in a compact vector space based on data associated with users and items (e.g., user-item interactions). Subsequently, a decoder, often a similarity metric, is employed to calculate similarity scores between user-item pairs based on their embeddings, followed by a loss function (e.g., MSE \cite{steck2013evaluation} and BPR \cite{rendle2012bpr}) tailored to different optimization goals and recommendation tasks.

Most current RSs focus on exploring various encoder architectures, evolving from early matrix factorization techniques \cite{he2016fast,baltrunas2011matrix,rendle2012bpr}, 
to Deep Neural Networks (DNNs) \cite{he2017neural,covington2016deep} and 
Graph Neural Networks (GNNs) \cite{wang2019neural,he2020lightgcn,sun2020disease,qu2021imgagn}.
However, the decoder, serving as the core component for calculating the similarity between users and items, has been overlooked. Most existing methods employ a uniform decoder, typically utilizing some common handcrafted metrics, such as the inner product \cite{he2020lightgcn,rendle2012bpr}, and L2 distance \cite{zhang2018metric,hsieh2017collaborative}, which may be suboptimal.
Here, we first conduct a preliminary experiment to validate that when we fix the other components of a recommender system, using different similarity metrics can lead to significant differences in the model's performance, which is illustrated in Figure \ref{fig:pe}. Based on this fact, 
some methods attempt to optimize the similarity metric. For example, NCF \cite{he2017neural} proposes to use a multi-layer perceptron (MLP) to replace the inner product, thereby enabling the learning of arbitrary non-linear interactions between users and items. However, the performance of such methods is sensitive to the architecture of the MLP. Another research line in this area is metric learning based methods, which utilize carefully designed variants of existing metrics (e.g., variants of the L2 distance \cite{park2018collaborative,tay2018latent})  or loss functions \cite{li2020symmetric,zhang2018metric}. However, such handcrafted metrics may not fully capture the diverse range of similarity patterns that can vary significantly across different domains. 



Inspired by recent advances in automated machine learning applied to RSs \cite{zheng2023automl,tang2023automl,qu2022single,chen2021learning,yin2024device,zheng2024personalized}, which can help free people from designing different components of RSs, the automated design of similarity metrics, however, remains an unexplored area. To bridge this gap, 
we propose an Automated Similarity Metric Generation method for recommendations, named AutoSMG, which can generate tailored similarity metrics from scratch. 
Specifically, we first construct a similarity metric space comprising a series of candidate metrics. To enable the construction of metrics from scratch, for each candidate metric, we sample from some of the most basic embedding operators, such as addition and normalization. These sampled operators are then assembled using a computational graph to represent the metric. After that, we utilize evolutionary algorithms, leveraging their parallel execution and global search capabilities  \cite{cabrera2016evolutionary}, to iteratively generate new candidate metrics and search for the best one. Furthermore, since fully training each candidate metric to evaluate its performance is time-consuming, we address this issue by adopting two strategies. One is an early stopping strategy, which leverages the early-phase performance of models on the validation set as a measure of the metric's fitness. The other is to train a surrogate model to predict the performance of candidate metrics. 


Overall, the contributions of this work are summarized as below:
\begin{itemize}
    \item To the best of our knowledge, this is the first work to introduce a similarity metric generation task in RSs, aiming to find the optimal metric that fits the other components of RSs.
    \item We introduce an Automated Similarity Metric Generation method, named AutoSMG, which efficiently searches for tailored similarity metrics from scratch using evolutionary algorithms. Moreover, our method is model-agnostic and can be integrated as a plugin into different recommendation model architectures.
    \item We validate our proposed method on public datasets across three different domains. The experimental results demonstrate the effectiveness of our method, which can outperform both commonly used handcrafted metrics and those generated by other search strategies.
\end{itemize}

The remainder of this paper is organized as follows: Section 2 reviews related work and identifies the research gap. Section 3 first formulates our problem and then provides a detailed description of our method. Section 4 presents our experimental setup and discusses our experimental results. Section 5 concludes the work and outlines future research directions.

\section{Related Work}
In this section, we will review relevant recommendation methods, including traditional RSs and the application of automated machine learning (AutoML) techniques to RSs.

\subsection{Recommender Systems}
As introduced in Figure \ref{fig:RSArchitecture}, the typical architecture of modern embedding-based RSs primarily consists of four components. Depending on the training data used, RSs can be divided into collaborative filtering methods \cite{he2016fast,rendle2012bpr,he2020lightgcn}, which rely solely on user-item interaction data, and content-based methods that leverage additional side information associated with users or items, such as users' social relationships \cite{fan2019graph,wu2018collaborative} and multimodal attributes of items \cite{tao2020mgat,wu2021mm}. The encoder serves as a function to map users and items into an embedding space, where much of the research in RSs focuses on designing various encoder architectures. This includes early methods based on matrix factorization \cite{he2016fast,baltrunas2011matrix,rendle2012bpr}. Subsequently, some approaches utilize deep learning techniques to capture complex nonlinear relationships between users and items by leveraging deep neural networks \cite{he2017neural,covington2016deep} and graph neural networks \cite{he2020lightgcn,fan2019graph,wu2022graph}. 
Depending on different recommendation tasks, various loss functions are selected as the optimization objective, common ones include Mean Squared Error (MSE) for regression tasks like rating prediction \cite{steck2013evaluation}, Bayesian Personalized Ranking (BPR) for ranking tasks such as Top-K recommendations \cite{rendle2012bpr}, and cross-entropy for classification tasks like CTR predictions \cite{guo2017deepfm}. However, there has been less exploration in decoders, namely the similarity metric, with existing methods mainly based on metric learning approaches, such as TransCF \cite{park2018collaborative} and LRML \cite{tay2018latent}, which introduce two carefully designed variants of the L2 distance, and other methods \cite{li2020symmetric} that jointly design combinations of metrics and loss functions. However, these methods are still manually designed, heavily relying on expert knowledge, and might not fully capture the diversity of similarity patterns that can vary significantly across different domains.

\subsection{AutoML in Recommender Systems}
To free people from designing different components of RSs, research in recent years has shifted from manual to automated design through the use of automated machine learning (AutoML) techniques \cite{zheng2023automl,tang2023automl}. 
For instance, some methods focus on the automated design of encoders.  NASR \cite{cheng2022towards} employs a greedy search strategy to find the optimal architecture for sequential recommendation tasks, while AutoCTR \cite{song2020towards} utilizes a multi-objective evolutionary algorithm to explore the two-level hierarchical search space.
Additionally, some methods \cite{joglekar2020neural,zhaok2021autoemb,zhao2021autodim,qu2022single} attempt to automatically design different user/item embedding sizes to reduce model size while maintaining performance. 
For content-based methods, they attempt automated feature selection \cite{wang2022autofield,luo2019autocross} and the design of various feature interaction strategies \cite{chen2019bayesian,liu2020autofis} to enhance the performance of models.
Moreover, there are approaches that automatically design loss functions. AutoLoss \cite{zhao2021autoloss} attempts to learn the weights of different loss combinations automatically. AutoLossGen \cite{li2022autolossgen} is a method similar to ours, aiming to generate loss functions from scratch using a set of basic mathematical symbols. However, directly applying this approach to similarity metric generation task presents challenges, primarily due to the differences in search spaces. Although these AutoML methods have shown promising performance in the automatic design of recommender systems, the design of decoders, specifically similarity metrics, remains largely unexplored. This identifies the gap our work aims to fill.

\begin{figure*}
    \centering
    \includegraphics[width=1\linewidth]{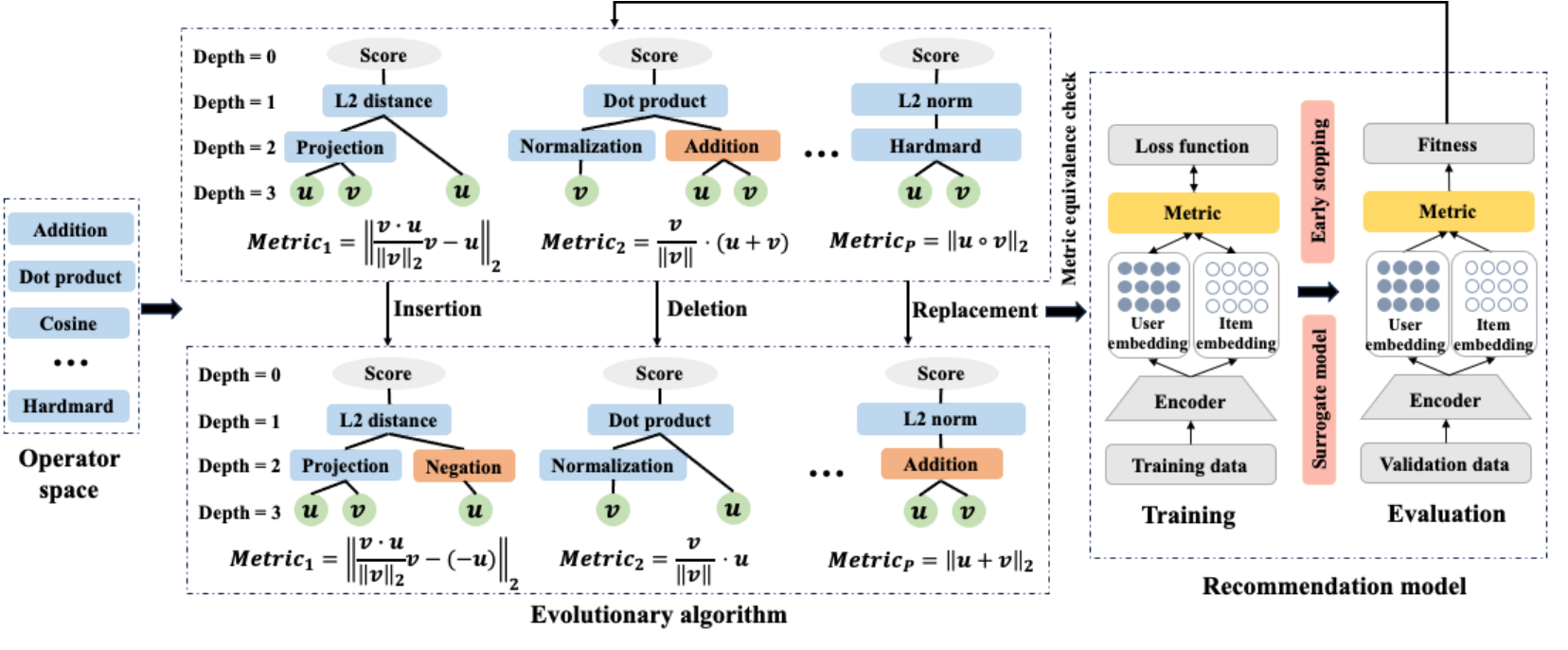}
    \caption{The architecture of the proposed AutoSMG.}
    \label{fig:autosmg}
\end{figure*}

\section{Proposed Method}
In this section, we first formulate the task of similarity metric generation for RSs, and then elaborate the proposed automated similarity metric generation method.

\subsection{Problem Formulation}
As we introduced earlier in Figure \ref{fig:RSArchitecture}, the recommender system is typically composed of the following four components:
\begin{itemize}
    \item \textbf{The user-item interaction data} $\mathcal{D}=\{\mathcal{U},\mathcal{I},\mathbf{A}\}$: It serves as the training data that represents users' preference. It includes a set of users $\mathcal{U}$, a set of items $\mathcal{I}$. The implicit interactions between users and items could be represented as an $|\mathcal{U}| \times |\mathcal{I}|$ adjacency matrix $\mathbf{A}$, where $a_{ui}=1 \in \mathbf{A}$ denotes there is an implicit interaction (e.g., click and purchase) between user $u \in \mathcal{U}$ and item $i \in \mathcal{I}$, otherwise $a_{ui}=0$.
    \item \textbf{The encoder} $f(\Theta)$: It represents a recommendation model parameterized by $\Theta$, used to learn the representations (i.e., embeddings) $\textbf{P} \in \mathbb{R}^{|\mathcal{U}| \times d}$ and $\textbf{Q} \in \mathbb{R}^{|\mathcal{I}| \times d}$ in a $d$-dimensional vector space, where $\textbf{p}_{u} \in \textbf{P}$ and $\textbf{q}_{i} \in \textbf{Q}$ denote the embeddings of user $u$ and item $i$, respectively.
    \item \textbf{The decoder (similarity metric) $SM(\textbf{p}_{u},\textbf{q}_{i})$}: It is used to calculate the similarity score between a pair of user and item based on their embeddings $\textbf{p}_{u}$ and $\textbf{q}_{i}$, thereby enabling recommendation services by suggesting the most similar items to users based on the ranking of similarity scores.
    \item \textbf{The loss function $\mathcal{L}$}: It is utilized to define the optimization goal based on various recommendation tasks. Here, we take the commonly used BPR (Bayesian Personalized Ranking) loss $\mathcal{L}_{BPR}$ \cite{rendle2012bpr} in Top-K recommendation tasks as an example \cite{he2016fast,he2020lightgcn}, which can be defined as follows:
    \begin{equation}
        \mathcal{L}_{BPR} = \sum_{(u,i,j) \in \mathcal{D}} \ln \sigma(SM(\textbf{p}_{u},\textbf{q}_{i})-SM(\textbf{p}_{u},\textbf{q}_{j}))
    \end{equation}
    where $i \in \mathcal{I}$ and $j \in \mathcal{I}$ represent items that have interacted with user $u \in \mathcal{U}$ (i.e., $a_{ui} = 1$) and items have not (i.e., $a_{uj}=0$), respectively.
\end{itemize}

Finally, the similarity metric search task for recommendation can be defined as finding the optimal similarity metric $SM^{*}(\cdot)$ for the given data $\mathcal{D}$, the encoder $f(\Theta)$, and the loss function $\mathcal{L}$ by the search algorithm $g(\cdot)$ as follows:
\begin{equation}
    SM^{*}(\cdot) = g(\mathcal{D},f(\Theta),\mathcal{L},\mathcal{C})
\end{equation}
where $\mathcal{C}=\{SM_{1}(\cdot),...\}$ represent the similarity metric space consisting of a set of candidate metrics.

\subsection{AutoSMG}

\begin{table*}[htbp]
    
  \centering
  \caption{The summary of the operator space.}
    \begin{tabular}{c|c|c|c|c|c}
    \toprule
    \multicolumn{2}{c|}{Vector Operations} & Expression & Arity & Input & Output \\
    \midrule
    \midrule
    \multicolumn{1}{c|}{\multirow{8}[2]{*}{Inter-vector operations}} & Addition & $\textbf{v}+\textbf{u} = (v_{1}+u_{1},...,v_{d}+u_{d})$       & 2     & vector & vector \\
          & Subtraction  & $\textbf{v}-\textbf{u} = (v_{1}-u_{1},...,v_{d}-u_{d})$        & 2     & vector & vector \\
          & Inner product & $\textbf{v}\cdot\textbf{u} = v_{1}u_{1}+...+v_{d}u_{d}$       & 2     & vector & scalar \\
          & Cosine similarity & $\frac{\textbf{v}\cdot\textbf{u}}{||\textbf{v}||||\textbf{u}||}$      & 2     & vector & scalar \\
          & Hardmard product & $\textbf{v}\circ\textbf{u} = (v_{1}u_{1},...,v_{d}u_{d})$       & 2     & vector & vector  \\
          & L1 distance & $||\textbf{v}-\textbf{u}||_{1} = |v_{1}-u_{1}|+...+|v_{d}-u_{d}|$       & 2     & vector & scalar  \\
          & L2 distance & $||\textbf{v}-\textbf{u}||_{2} = \sqrt{(v_{1}-u_{1})^{2}+...+(v_{d}-u_{d})^{2}}$       & 2     & vector & scalar  \\
          & Projection &  $proj_{\textbf{u}}\textbf{v} = (\frac{\textbf{u}\cdot\textbf{v}}{||\textbf{u}||_{2}})\textbf{u}$     & 2     & vector & vector  \\
    \midrule
    \multicolumn{1}{c|}{\multirow{5}[2]{*}{Intra-vector operations}} & L1 norm & $||\textbf{u}||_{1} = |u_{1}|+...+|u_{d}|$      & 1     & vector & scalar  \\
          & L2 norm &  $||\textbf{u}||_{2} = \sqrt{u_{1}^{2}+...+u_{d}^{2}}$     & 1     & vector & scalar  \\
          & Normalization &  $\frac{\textbf{u}}{||\textbf{u}||}$     & 1     & vector & vector \\
          & Scalar multiplication &  $c\textbf{u} = (cu_{1},...,cu_{d})$     & 1     & vector & vector \\
          & Negation &  $-\textbf{u} = (-u_{1},...,-u_{d})$     & 1     & vector & vector \\
    \bottomrule
    \end{tabular}%
  \label{tab:search space}%
\end{table*}%

In this work, we decompose the similarity metric generation task into two sub-tasks: one is to construct a similarity metric space that can include a series of candidate metrics, and the other is to efficiently search within the metric space to find the optimal one. To this end, we propose an automated similarity metric generation method for recommendation, named AutoSMG. The architecture of the method is illustrated in Figure \ref{fig:autosmg}. Specifically, in the metric generation phase, we first select a series of basic embedding operators to construct the operator space. Then, we randomly sample within the operator space and use computational graphs to represent computable candidate metrics, thereby constructing a metric space. Secondly, in the metric search phase, to find the optimal metric among the candidates, we employ an evolutionary algorithm to search and generate new metrics within the metric space. Since fully training each candidate metric to evaluate its performance is time-consuming, we address this issue by adopting two strategies. One is an early stopping strategy that uses the performance of the models in the validation set in the early training phase as the fitness of the metric. The other is to train a surrogate model to predict the performance of candidate metrics. The detailed algorithmic process is presented in Algorithm \ref{alg_ours}.

\subsubsection{Metric generation}
Since the existing similarity metrics, such as inner product and L2 distance, are limited, they cannot guarantee to meet diverse similarity patterns for different domains. Therefore, the metric generation phase aims to explore a vast space of potential similarity metrics beyond the limitations of pre-defined ones, offering a tailored solution that can better meet the specific requirements of different recommendation models or datasets. To achieve this, we choose to construct the operator space from some of the most basic embedding operators and then sample a set of operators from it, using computational graphs to represent candidate metrics. The details are as follows:

\textbf{Operator Space:} In the context of RSs, the calculation of similarity scores typically occurs between embeddings, i.e., the user embedding and item embedding. Therefore, we construct the operator space based on a series of most basic embedding operators, which have been summarized in Table \ref{tab:search space}. Specifically, we categorize these operators into two types including (1) inter-vector operators involve operations between two embeddings, with an arity of 2. These operators are designed to measure the relationship between pairs of embeddings. (2) intra-vector operators involve operations on a single embedding itself, with an arity of 1. These operators are used to manipulate or derive new properties from individual embeddings. Moreover, to ensure that the computational graphs can generate computable metrics, we define the input and output types for each operator, ensuring compatibility between operators in the computational graph.

\textbf{Metric Representation:} Each candidate metric is represented by a computational graph by sampling a set of operators from the operator space. Specifically, as shown in Figure \ref{fig:autosmg}, the root node (depth=0) represents the output (i.e., the similarity score).
The leaf nodes, i.e., the nodes at the last layer (depth=3), are inputs sampled from $\{\textbf{u},\textbf{v},\textbf{1}\}$ including the user embedding $\textbf{u}$, item embedding $\textbf{v}$, and an additional $\textbf{1}$ vector to enhance flexibility. 
The computation nodes in the intermediate layers are obtained by randomly sampling from the operator space. 
Notably, to ensure the computability of the computational graph, it is necessary to align the input and output types between computation nodes. For example, since the root node at depth 0 represents the similarity score, which is a scalar, only those operators that output a scalar type, such as L2 distance, can be selected for sampling at the first layer.
Furthermore, at the last layer's leaf nodes, sampling is done without replacement to ensure that the computation process includes both user embedding and item embedding. 
Only after both user embedding and item embedding have been selected and removed from the sampling pool $\{\textbf{u},\textbf{v},\textbf{1}\}$, will they be put back for resampling.

\subsubsection{Metric search}
Given the parallelizability and global search capabilities of evolutionary algorithms, which have been validated in many AutoML tasks \cite{liang2019evolutionary,real2020automl}, we employ evolutionary algorithms to iteratively search for and generate new metrics. Specifically, the evolutionary search process includes the following steps:

\textbf{Population Initialization}: At the beginning, we randomly initialize a population by generating $N$ candidate metrics, denoted as $\mathcal{M} = \{SM_{1}(\cdot),\dots,SM_{n}(\cdot)\}_{n=1}^{N}$. Additionally, to avoid repetitively evaluating mathematically equivalent similarity metrics, we conduct a Metric-Equivalence Check for every pair of metrics of the population. Specifically, we randomly initialize a set of user-item embedding pairs $\mathcal{D}_{MEC}=\{\textbf{u}_{k},\textbf{v}_{k}\}_{k=1}^{K}$, then calculate similarity score vectors of $N$ metrics in the population $\mathcal{M}$ on $\mathcal{D}_{MEC}$, denoted as $\textbf{S}=\{\textbf{s}_{1},...,\textbf{s}_{n}\}_{n=1}^{N}$, where $\textbf{s}_{n} \in \mathbb{R}^{K}$ represents similarity score vector of the $n$-th metric $SM_{n}(\cdot)$ on $\mathcal{D}_{MEC}$. The equivalence between every two metrics in $\mathcal{M}$, e.g., $SM_{a}(\cdot), SM_{b}(\cdot) \in \mathcal{M}$, is assessed by comparing their similarity score vectors $\textbf{s}_{a}$ and $\textbf{s}_{b}$ as below:

\begin{equation}
\text{Equivalent}(SM_{a}(\cdot), SM_{b}(\cdot)) = 
\begin{cases} 
\text{True} & \text{if } |\textbf{s}_{ak} - \textbf{s}_{bk}| < \delta \text{ for all } k \\
\text{False} & \text{otherwise}
\end{cases}
\label{equ:mqc}
\end{equation}
where $\delta$ is a predefined threshold. If two metrics are determined to be equivalent, indicating that they are mathematically equivalent or very similar in their evaluation of the user-item pairs. One of them will be randomly removed, and a new metric will be randomly generated to replace it. 

\textbf{Metric Evaluation:} To evaluate the fitness of each candidate metric in the population, theoretically, it should be fully trained on the given encoder and training dataset, and then its performance (e.g., the NDCG@20 score) on the validation set can serve as its fitness, denoted as $\textbf{y} = \{y_{1},...,y_{n}\}_{n=1}^{N}$. However, this process is time-consuming. To address this issue, we adopt two fitness approximation strategies, including early stopping strategy and surrogate model strategy, to obtain the predicted metric fitness, denoted as $\hat{\textbf{y}} = \{\hat{y}_{1},...,\hat{y}_{n}\}_{n=1}^{N}$.
\begin{itemize}
    \item Early Stopping Strategy (ES): This strategy involves training the model with the candidate metric for a limited number of epochs rather than completing the full training cycle. The idea is to estimate the metric's performance early in the training process, which significantly reduces the computational cost. The performance of the metric on the validation set after these initial epochs is used as a proxy for its fitness.
    \item Surrogate Model Strategy (SUR): This approach involves using a surrogate model $h(\Theta_{sur})$, which is parameterized by $\Theta_{sur}$, to predict the fitness of a candidate metric without actually training it. 
    This surrogate model is trained on the dataset $\mathcal{D}_{SUR}=\{(SM_{z}(\cdot),y_{z})\}_{z=1}^{Z}$ consisting of metrics $SM_{z}(\cdot)$ whose fitness $y_{z}$ has been previously evaluated (after full training). In this way, the surrogate model can quickly predict the fitness of new candidate metrics without the need for extensive training. 
    
    Specifically, this work uses a sequence-to-one surrogate model architecture, which is well-suited for processing the sequential nature of the computational graphs representing candidate metrics, and each operator within the operator space is assigned with a unique embedding. In this way, the computational graph of a candidate metric is transformed into a sequence of operator embeddings according to the graph's topology. This sequence represents the flow of operations in the metric calculation and serves as the input to the surrogate model. The surrogate model is then optimized by comparing the predicted fitness value $\hat{y}_{z} = h(\Theta_{sur},SM_{z}(\cdot))$ against the actual fitness values $y_{z}$ via the Mean Squared Error (MSE) loss as below:
    \begin{equation}
        \mathcal{L}_{MSE} = \frac{1}{|Z|}\sum_{(SM_{z}(\cdot),y_{z}) \in \mathcal{D}_{SUR}}^{Z} (\hat{y}_{z} - y_{z})^{2}
    \end{equation}
\end{itemize}

\textbf{New Metric Generation:}
After obtaining the fitness of each candidate metric in the current generation, we randomly select a proportion of $\gamma$ from the current population to generate new metrics, where $\gamma$ is the mutation ratio to balance exploration (searching for new potential metrics) and exploitation (refining existing metrics).
Specifically, as shown in Figure \ref{fig:autosmg}, we employ the following three types of mutation operations \cite{li2022autoloss}:
\begin{itemize}
    \item Insertion: The insertion operation involves randomly selecting a non-root node and its parent node in the computational graph and inserting an operator between them. Notably, the choice of the operator to insert is based on the computational
    constraints, ensuring that the new operator is compatible with the types of its parent and child nodes in the graph.
    \item Deletion:  The deletion operation randomly removes an intermediate computational node from the graph, subsequently selecting a child node that satisfies computational constraints to replace the deleted node as the new child of its parent. If both children are eligible, one is chosen randomly; if neither qualifies, the deletion attempt is redirected to a different node. 
    \item Replacement: The replacement mutation operation strategically swaps a non-root operator in the current metric's computational graph with another operator randomly selected from the operator space. If the new operator's arity matches the replaced one, the child nodes remain untouched; If it's higher, an additional child node is introduced to accommodate the extra input; If it's lower, a child node is randomly removed to align with the reduced arity.
\end{itemize}

\textbf{Selection:}
After generating new metrics, we first apply the Metric-Equivalence Check to identify and eliminate mathematically equivalent metrics. Following this, the fitness of the new metrics is evaluated based on strategies outlined in Metric Evaluation. After that, only the Top-$N$ metrics, where $N$ is the predefined population size, are retained for the next generation. This selection process prioritizes the metrics with the highest fitness scores, thereby concentrating the population on the most promising solutions and continuously refining the quality of metrics across generations.

\begin{algorithm}[!ht]
  \renewcommand{\algorithmicrequire}{\textbf{Input:}}
  \renewcommand{\algorithmicensure}{\textbf{Output:}}
  \caption{AutoSMG} \label{alg_ours}
  \begin{algorithmic}[1]
    \Require Dataset $\mathcal{D}$; Encoder $f(\Theta)$; Loss function $\mathcal{L}$, \dots
    \Ensure Optimal decoder (i.e., similarity metric) $SM^{*}(\cdot)$
    \State  Initialize user embedding $\textbf{P}$, item embedding $\textbf{Q}$, and model parameters $\Theta$
    \State Initialize population $\mathcal{M}^{(0)} = \{SM_{1}^{(0)}(\cdot),\dots,SM_{n}^{(0)}(\cdot)\}_{n=1}^{N}$.
    \State Metric-Equivalence Check by Equation (\ref{equ:mqc}) 
    \State Metric evaluation $\hat{\textbf{y}}^{(0)} = \{\hat{y}_{1}^{(0)},...,\hat{y}_{n}^{(0)}\}_{n=1}^{N}$ \Comment{ES or SUR}
    \For {each round t =0, ..., $T$} \Comment{Evolution generations}
        \State Generate $\gamma N$ new metrics $\hat{\mathcal{M}}^{(t)} = \{\hat{SM}_{n}^{(t)}(\cdot)\}_{n=1}^{\gamma N}$.
        \State Metric-Equivalence Check for generated $\hat{\mathcal{M}}^{(t)}$
        \For{$\hat{SM}_{n}^{(t)}(\cdot)\in \hat{\mathcal{M}}^{(t)}$ \textbf{in parallel}}
        \State Metric evaluation $\bar{\textbf{y}}^{(0)} = \{\bar{y}_{1}^{t)},...,\bar{y}_{n}^{(t)}\}_{n=1}^{\gamma N}$ \Comment{ES or SUR}
        \EndFor
      \State Select Top-$N$ metrics as new generation population $\mathcal{M}^{(t+1)}$
    \EndFor

    \end{algorithmic}
\end{algorithm}

\begin{table*}[!htbp]
  \centering
  \caption{The performance of different metrics on Tok-K recommendation task.}

    \begin{tabular}{c|l|cc|cc|cc}
    \toprule
    \multicolumn{2}{c|}{\multirow{2}[4]{*}{Method}} & \multicolumn{2}{c|}{MovieLens-1M} & \multicolumn{2}{c|}{Gowalla} & \multicolumn{2}{c}{LastFM} \\
\cmidrule{3-8}    \multicolumn{2}{c|}{} & Recall@20 & NDCG@20 & Recall@20 & NDCG@20 & Recall@20 & NDCG@20 \\
    \midrule
    \midrule
    \multirow{7}[2]{*}{NCF} & Inner Product & 0.2071 & 0.3059 &0.1341&0.1051&0.0589&0.0392\\
          & MLP   & 0.1912 & 0.2998 & 0.1324 & 0.1008 & 0.0483 & 0.0276\\
          & TransCF & 0.1909  & 0.2913  & 0.1294  & 0.0975  & 0.0456 & 0.0248  \\
          & Random search  & 0.1921 & 0.3078 & 0.1342 &0.0916&0.0661&0.0409\\
          & AutoLossGen & 0.2045 & 0.3030  & 0.1338  & 0.0902  & 0.0671 & 0.0425  \\
          & AutoSMG-ES & \textbf{0.2323} & \textbf{0.3258} & \textbf{0.1511}& \textbf{0.1201}&  0.1068 &0.0667\\
          & AutoSMG-SUR & 0.2213  & 0.3147 & 0.1502 & 0.1183  & \textbf{0.1097} & \textbf{0.0682}  \\
    \midrule
    \multirow{7}[2]{*}{LightGCN} & Inner Product & 0.2437 & 0.3631 &0.1814&0.1540&0.2680&0.2096\\
          & MLP   & 0.2257& 0.3417 &0.1742&0.1465&0.2531&0.1833\\
          & TransCF & 0.2136 & 0.3335 & 0.1688  & 0.1394 & 0.2503 & 0.1774  \\
          & Random search  & 0.2066& 0.3303&0.1775&0.1488&0.2323&0.1801\\
          & AutoLossGen &  0.2535  & 0.3756  & 0.182  & 0.1543   & 0.2676  & 0.2054  \\
          & AutoSMG-ES & \textbf{0.2677}& \textbf{0.3965}&\textbf{0.1826}&\textbf{0.1555}&0.2781&0.2187\\
          & AutoSMG-SUR & 0.2599  & 0.3812  & 0.1822  & 0.1547  & \textbf{0.2784} & \textbf{0.2196}  \\
    \bottomrule
    \end{tabular}%
\label{tab:Top-K}%
\end{table*}%

\begin{table*}[!htbp]
  \centering
  \caption{Top-3 optimal metrics of each encoder-dataset pair.}

    \begin{tabular}{l|c|c|c}
    \midrule
Metrics& Top1 & Top2  &Top3\\
    \midrule
    \midrule
    NCF-MovieLens & $(\textbf{v}+\textbf{1})\cdot (2\frac{\textbf{u}}{||\textbf{u}||})$& $2\frac{\textbf{u}}{||\textbf{u}||}+\textbf{v}$&$\textbf{u}\cdot(\textbf{v}+c\textbf{1})$\\
    NCF-LastFM& $-\frac{(\textbf{v}-\textbf{1})\cdot\frac{\textbf{u}}{||\textbf{u}||}}{||\textbf{v}-1||_{2}}(\textbf{v}-\textbf{1})$& $(\textbf{u}-\textbf{1})\cdot(\frac{(\textbf{v}-\textbf{1})\cdot\frac{\textbf{u}}{||\textbf{u}||}}{||\textbf{v}-1||_{2}})(\textbf{v}-\textbf{1})$&$(2\frac{\textbf{u}}{||\textbf{u}||})\cdot (\textbf{v}+c\textbf{1})$\\
 NCF-Gowalla& $\frac{c\textbf{u}}{||c\textbf{u}||}\cdot (2\textbf{v}+\textbf{1})$& $(\textbf{v}+\textbf{1})\cdot (\frac{\textbf{u}}{||\textbf{u}||}+\frac{\textbf{v}}{||\textbf{v}||})$&$(2\frac{\textbf{u}}{||\textbf{u}||})\cdot (\textbf{v}+\textbf{1})$\\
    \midrule
    LightGCN-MovieLens
& $\frac{c\textbf{u}}{||c\textbf{u}||}\cdot(\frac{\textbf{v}\cdot\textbf{v}}{||\textbf{v}||_{2}}\textbf{v}+\textbf{v})$& $\frac{c\textbf{u}}{||c\textbf{u}||}\cdot (2\textbf{v})$&$\frac{\textbf{u}}{||\textbf{u}||}\cdot (2\textbf{v}-\textbf{u})$\\
    LightGCN-LastFM
& $\sum(\frac{\textbf{u}\cdot\textbf{v}}{||\textbf{u}||_{2}}\textbf{u})$& $\sum{}(2\textbf{u}-\frac{\textbf{u}\cdot\textbf{v}}{||\textbf{u}||_{2}}\textbf{u})$&$\sum(\textbf{u}\otimes\textbf{v}-\frac{\textbf{1}}{||\textbf{1}||})$\\
    LightGCN-Gowalla& $(\frac{\textbf{u}}{||\textbf{u}||}+c\textbf{u})\cdot(\frac{\textbf{u}\cdot\textbf{v}}{||\textbf{u}||_{2}}\textbf{u})$& $(\frac{\textbf{u}}{||\textbf{u}||}+c\textbf{u})\cdot\textbf{v}$&$c(\frac{\textbf{u}\cdot\textbf{u}}{||\textbf{u}||_{2}}\textbf{u}) \cdot \textbf{v}$\\
    \bottomrule
    \end{tabular}%
  \label{tab:metrics}%
\end{table*}%

\section{Experiment}

In this section, we conduct experiments to answer the following research questions (RQs):
\begin{itemize}
    \item RQ1: How does the performance of our automatically generated metrics compare to that of handcrafted metrics and other metric generation methods?
    \item RQ2: Is our method applicable to different recommendation model architectures?
    \item RQ3: How do different components of the proposed method impact model performance?
    \item RQ4: How do different hyperparameters impact model performance?
\end{itemize}

\subsection{Dataset}

We validate our method on three commonly used recommendation system datasets from different domains, including MovieLens-1M \cite{harper2015movielens}, which comprises user ratings for movies; Gowalla \cite{liang2016modeling}, featuring user check-ins; and the LastFM \cite{wang2019kgat} music listening dataset. To ensure a fair comparison, we follow the data processing used by LightGCN\footnote{https://github.com/gusye1234/LightGCN-PyTorch}. The statistical information of the datasets is summarized in Table \ref{tab:datasets} below. 

\begin{table}[!htbp]
  \centering
  \caption{Statistics of datasets}

    \begin{tabular}{c|c|c|c|c}
    \toprule
    Dataset & \# Users & \# Items &  \# Interactions&Density \\
    \midrule
    \midrule
    MovieLens-1M& 6,040&3,706&  1,000,210&0.04468 \\
    Gowalla & 29,858&  40,891&  1,027,370&0.00084 \\
    LastFM & 1,892&  4,489& 52,668&0.00620 \\
    \bottomrule
    \end{tabular}
\label{tab:datasets}
\end{table}

\subsection{Baseline}
To evaluate the proposed automated similarity metric generation method, we compare it against two categories of methods: handcrafted metrics and automatically generated metrics as below:

\begin{itemize}
    \item \textbf{Handcrafted metrics:}
    \begin{itemize}
        \item \textbf{Inner product:} A widely used metric that calculates the directional similarity between two embeddings. It is simple yet effective for many recommendation tasks.
        \item \textbf{MLP:} A method that utilizes a MLP to capture the complex non-linear relationships between users and items.
        \item \textbf{TransCF \cite{park2018collaborative}:} A metric learning-based method that designs a improved variant of the L2 distance, incorporating translation vectors to measure the distance between embeddings.
    \end{itemize}
    \item \textbf{Automatically generated metrics:}
    \begin{itemize}
        \item \textbf{Random search:} A baseline method that conducts an automated search by randomly selecting an equivalent number of candidate metrics from the metric space, comparable to our method. This strategy assesses the efficiency of unguided exploration within the metric space, questioning the need for directed optimization.
        \item \textbf{AutoLossGen \cite{li2022autolossgen}:} As there are no existing algorithms specifically tailored for metric generation, we adapt AutoLossGen, an algorithm originally designed for loss function generation using reinforcement learning, by replacing its search space with our metric space. 
    \end{itemize}
\end{itemize}

\subsection{Setup}

For the encoder, we select two state-of-the-art (SOTA) architectures from mainstream methodologies: NCF \cite{he2017neural}, which is based on deep neural networks and matrix factorization, and LightGCN \cite{he2020lightgcn}, which is based on graph neural networks. We implement these two methods based on the code provided by the authors and merely replace their original metrics with our generated metrics for comparison. The embedding dimension for users and items is uniformly set to 64. The population size for the evolutionary algorithm is set to 50, and the number of generations for the search is set to 100. For the early stopping strategy, we chose a training duration of 10 epochs. The BPR is selected as the loss function throughout our experiments. For the surrogate model, we adopted an LSTM structure as the backbone. We use two widely utilized evaluation metrics for recommender systems, recall@20 and ndcg@20, for Top-K recommendation tasks. The mutation ratio $\gamma$ is set to $0.7$. Moreover, the candidate metrics' performance on the validation set in terms of NDCG@20 is used as the fitness to select metrics.

\subsection{Top-K Recommendation (RQ1,RQ2)}

To validate the effectiveness of our proposed method, we compare its performance on the Top-K recommendation task against other baseline methods. Moreover, to validate that our framework is model-agnostic, which can be seamlessly integrated into different recommendation model frameworks, we integrated the proposed methods into the NCF and LightGCN methods, respectively. In addition, AutoSMG-ES and AutoSMG-SUR represent our method with the early stopping strategy and the surrogate model strategy, respectively. In Table \ref{tab:Top-K}, we report the performance of different methods on three datasets in terms of Recall@20 and NDCG@20. In Table \ref{tab:metrics}, we present the top-3 metrics generated by our method for each given (encoder-dataset) pair. From the experimental results, we can observe:

\begin{itemize}
    \item Methods based on automatically generated metrics outperform those based on handcrafted metrics in most cases. This validates the necessity of our research problem: a tailor-made metric can effectively enhance the performance of recommender systems.
    \item Within the realm of handcrafted metrics-based methods, the inner product method surpasses the MLP method and metric learning-based method in most cases. A possible reason, as previously mentioned, is the MLP method's sensitivity to its own architecture.
    \item In automatically generated metrics-based methods, our approach exceeds the performance of Random search and AutoLossGen. This success is attributed to the global search capability of evolutionary algorithms, in contrast to the AutoLossGen method, which, based on reinforcement learning, might get trapped in local optima.
    \item Our method achieves optimal results on both the NCF, which is based on deep matrix factorization, and LightGCN, which is based on graph neural networks, proving our method's model-agnostic characteristic.
    \item The variant based on a surrogate model, AutoSMG-SUR, performs better on the LastFM dataset than the variant based on an early stopping strategy, AutoSMG-ES. Conversely, AutoSMG-ES performs better on the other two datasets. A possible reason is that the surrogate model trains more effectively on the LastFM dataset, enabling it to predict the performance of metrics more accurately.
    \item For a given encoder-dataset pair, the best-performing metrics share similar structures. For instance, the Top-1 and Top-2 metrics for the (LightGCN-MovieLens-1M) pair, as well as the Top-2 and Top-3 metrics for the (NCF-LastFM) pair, exhibit structural similarities. This observation suggests that certain structural elements within the metrics may be particularly effective for specific types of data or model architectures. It highlights the potential for identifying underlying patterns or components that contribute to metric performance.
\end{itemize}

\begin{table}[!htbp]
  \centering
  \caption{Ablation study results on MovieLens-1M.}

    \begin{tabular}{l|c|c}
    \toprule
    Method & Speed-up & NDCG@20\\
    \midrule
    \midrule
    Vanilla Evolution & 1x  &   0.398 \\
    w/ ES  &  10x &             0.3788 \\
    w/ SUR &  52x &             0.3764 \\
    w/ ES+MEC &   12x &         0.3965 \\
    w/ SUR+MEC &   56x &        0.3812 \\
    \bottomrule
    \end{tabular}%
  \label{tab:AB}%
\end{table}%

\begin{figure*}[!ht]
\centering
\includegraphics[width=1\textwidth]{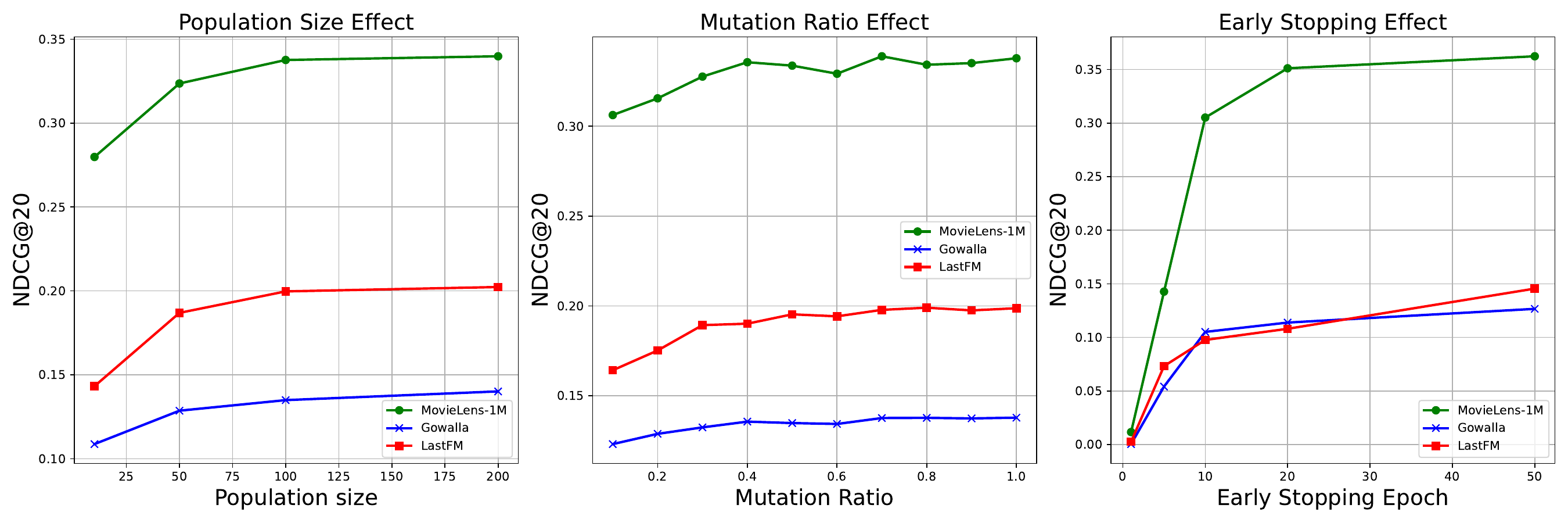} 
\caption{The model performance with various hyperparameters.}
\label{fig:HYPA}
\end{figure*}

\subsection{Ablation Study (RQ3)}
To investigate the impact of different fitness approximation strategies on the efficiency of model search, we initially implement a method without any approximation strategy, denoted as vanilla evolution. Subsequently, we implement variations of the method based on the early stopping strategy, denoted as \textit{w/ ES}, and on the surrogate model strategy, denoted as \textit{w/ SUR}. Moreover, we integrate the metric equivalence check (MEC) into these strategies to further refine the search process. We use LightGCN as the encoder, and report the speed-up ratio of search and NDCG@20 on the MovieLens-1M. Other settings are the same as those in the above Top-K recommendation experiment. The experimental results are presented in the table \ref{tab:AB}, allowing us to observe:
\begin{itemize}
    \item The two approximation strategies have proven to be effective in accelerating the model search process. This outcome is logical given that the vanilla evolution approach requires complete training for each candidate metric, which is time-consuming.
    \item The SUR method is faster than the ES method. The reason behind this is that SUR requires a one-time full training upfront to develop the surrogate model. Once this model is trained, it can predict the performance of new metrics without the need for additional extensive training. 
    \item The integration of MEC further accelerates the search process, indicating the presence of many mathematically equivalent metrics within the search space.
\end{itemize}

\subsection{Hyperparameter Analysis (RQ4)} 

To explore the impact of different hyperparameters on the model's performance, we conducted hyperparameter analysis experiments focusing on three key hyperparameters: 1. The impact of the population size N, for which we set N=[10,50,100,200]; 2. The influence of the mutation ratio $\gamma$, which we varied from 0.1 to 1 with stepsize 0.1; 3. The effect of the stopping epoch in the early stop strategy, for which we set the stopping epochs at 1, 5, 10, 20, 50. We used LightGCN as the encoder, with other settings consistent with those of the Top-K recommendation experiments. We report the NDCG@20 metric on the three datasets in Figure \ref{fig:HYPA}. From the experimental results, we observe that
\begin{itemize}
    \item As the population size increases, the model's performance initially improves significantly, indicating that even a limited increase in the diversity of the population can enhance the exploration capabilities of the algorithm significantly. 
    However, as the population size continues to grow, the benefit tapers off, likely due to the algorithm reaching a point where additional diversity doesn't contribute as much to finding better solutions. This implies there's an optimal range for population size that balances computational efficiency with performance gains.
    \item  The mutation ratio is crucial in balancing exploration and exploitation. A lower selection ratio indicates that fewer metrics are selected for the next generation, which may lead to rapid convergence but also risks missing out on potentially superior metrics. Conversely, a higher selection ratio allows for more extensive exploration but can slow down the convergence. 
    The optimal ratio found around 0.7 suggests that allowing a substantial proportion of the population to breed strikes a balance where the algorithm can sufficiently explore the metric space while still focusing on the most promising solutions.
    \item The finding that a very low early stop epoch results in poor model performance is intuitive. Early stopping in this context might prevent the model from adequately learning and approximating the true performance of the metrics, leading to suboptimal selection and propagation of solutions. 
    On the other hand, as the stopping epoch increases, the model has more time to train and therefore can more accurately estimate the performance of different metrics. The stabilization of results with longer epochs indicates that the models reach a level of training sufficient to capture the quality of the metrics.
\end{itemize}

\section{Conclusion}

In this work, we introduce an unexplored task in the realm of recommender systems: the similarity metric generation task, which aims to identify the optimal similarity metric for a given dataset, encoder, and loss function. To address this challenge, we propose an Automated Similarity Generation method for recommendations, named AutoSMG. Specifically, we first introduce an operator space and then combine it with computational graphs to construct a metric space. Subsequently, we utilize an evolutionary algorithm to iteratively search for the optimal metric within this metric space. Moreover, we propose using early stopping strategies and surrogate models to approximate the performance of candidate metrics, thereby accelerating the search process. Experiments conducted on three publicly available datasets from different domains demonstrate that our method outperforms handcrafted metrics as well as other automated metric generation algorithms. Additionally, our method can be seamlessly integrated as a plugin into various recommendation architectures. In the future, we aim to apply our method to other recommendation tasks and explore adaptive metrics to accommodate dynamic changes in data distribution. Additionally, we also plan to jointly search for metrics and loss functions, considering their combined effect on the user-item similarity patterns.



\bibliographystyle{ACM-Reference-Format}
\bibliography{main}


\end{document}